# DOPPLER-LIKE EFFECT

# AND DOUBTFUL EXPANSION OF UNIVERSE

Edward Szaraniec

Cracow University of Technology, ul. Warszawska 24, 31-155 Kraków, Poland

edszar@usk.pk.edu.pl

**Abstract**

The distance contraction, as observed in electrical soundings over horizontally stratified earth (static system), is identified as a counterpart of Doppler shift in dynamical systems.

Identification of Doppler-like effect in a stock-still systems makes it possible to give an alternative answer to the question about an effective cause of the Doppler shift, which sounds: the inhomogeneities. This answer opens different static as well as kinematic possibilities, which challenge established theories of expanding universe and energizing big bang.

The energy propagating in stratified universe of layers exhibits a shift which could be attributed not only to the expansion (Hubble's theory) but alternatively to fluctuations in material properties (inhomogeneities).

## A. Introduction

*Background.* In electrical soundings over horizontally stratified earth the equivalence of differently layered media is known since early days of the method [1]. The effect of resistivity sounding at the surface of a massive layering is the same as that for a finely layered structure provided that depth scale is stretched at each point by the local value of the so-called pseudo-anisotropy coefficient and, in addition, the resistivity is specified as a kind of average resistivity [2,3]. This effect of equivalence has been confirmed in a multitude of geophysical cases.

In geoelectric prospecting over horizontal layering it is up to standard to take this effect into account. Its exposition is a firm subject of geoelectrical monographs, and textbooks.

*Rationale.* In geophysical investigation of horizontally stratified earth the inhomogeneous media are studied under different scales. There are at least two scales involved: 1) the scale of inhomogeneities (microscale), 2) less detailed scale (macroscale) related with the size of a region under investigation.

Horizontal stratification implies material property (proper impedance) varying in one particular direction. In perpendicular directions the proper impedance being constant, this aspect of directional nonuniformity finds its expression in geophysical notion of pseudoanisotropy.

Of particular interest is a class of finely layered media exhibiting a repetitive or cyclic structure (proper impedance evolving between high and low values). Following geophysical theory, such pseudoanisotropic media are reparametrized under cumulative transformation, to get a massive layering. The rationale for use of these transformations is based on the knowledge that for cyclic media the transformed impedance is specified as pseudoimpedance (kind of mean impedance), and the scale is stretched by an overall value of the so-called pseudoanisotropy coefficient.

For a given cyclic pseudoanisotropic medium there exists a homogeneous isotropic medium that behaves, in the limit, exactly as does the given medium under the same excitation. This is to mean that a medium, which is viewed in the microscale as cyclic pseudoanisotropic, appears as a homogeneous medium when viewed in a macroscale. This is accompanied by distance contraction.

*Physical basis.* Geophysical static theory of cyclic stratified media [2,4,5,6] is generalized or modified in this letter in various aspects. In substance, the framework is completed by time domain, and the medium is considered to have a repetitive (cyclic) structure in all directions.

It is worth noting that the formulation developed for horizontally stratified (one dimensional) media is applicable to general situations when proper impedance vary in space (three dimensions). To this end the medium is assumed to be discretized in all directions, and the results, having no side effects and relative to a facultative direction, could be generalized to all the directions.

If medium viewed in a macroscale exhibits the pseudoimpedance and contraction coefficient which are both constant over a region of interest, it is said to be pseudohomogeneous. The rationale for studying such media is a multitude of geophysical observations transportable to 3-D space. The pseudohomogeneity in 3-D approach might be an inherent property of the media of physical interest, provided that the micro- and macro- scales are sufficiently different. This expectation is based on cumulative nature of the transformation involved as well as on isotropy postulated for largest regions.

### B.  Governing equation

To conceptualize the problem, the governing equation is taken in the form of scalar wave equation. For a time harmonic field U, we have

$$\rho \nabla \cdot \rho^{-1} \nabla U(r,t) + k^2 U(r,t) = 0 \tag{1}$$

when $U(r,t)$ can be either $U_x$, $U_y$, or $U_z$ field component, and $k = \omega^*/v$.

The above is the Helmholtz wave equation for an inhomogeneous, isotropic, and source-free medium. In addition, proper impedance, $\rho(r)$, and wave velocity, $v(r)$, are both functions of position, $r = r(x,y,z)$.

Following classical theory, in a stock-still system there is no anomalous frequency, that is to say $\omega^*$ is the same as wave frequency $\omega$ ($\omega^* = \omega$). When Doppler shift, $\omega^* = \omega H$, is observed, we may define $v^* = v/H$, that is to say $k = \omega/v^*$. In this case $v^*$ takes interpretation in terms of moving source and/or expanding medium. Alternative interpretation of anomalous frequency is just the point of this letter.

### C. Pseudohomogeneity

*Reparametrization.* We consider stratified medium, locally (in microscale) materialized through proper impedance varying possibly in all directions, but we are considering variations in only one direction, that is to say $\rho(r)$. Let a larger region of size W be investigated by an observer located at the origin, where the cumulative parameters are introduced:

$$T(r) = \int_0^w \rho(r)\,dr, \quad \text{and} \quad S(r) = \int_0^w \rho^{-1}(r)\,dr, \quad w \,\varepsilon\, W \tag{2}$$

These are used to define pseudoimpedance, and pseudodistance:

$$\rho^* = \sqrt{T/S}, \quad \text{and} \quad r^* = \sqrt{T \cdot S} \tag{3}$$

as well as pseudoanisotropy coefficient

$$\Lambda = r^*/r, \quad \Lambda \geq 1 \tag{4}$$

The bound for $\Lambda$ is known in mathematical geophysics[2]. For further use the contraction coefficient is introduced

$$\theta = \Lambda - 1, \quad \theta \geq 0 \tag{5}$$

These way the medium is reparametrized into pseudoimpedance varying in only one pseudodirection, $\rho^*(r^*)$.

*Cyclic, and pseudohomogeneous medium.* When proper impedance of a medium is varying between high and low values, in a direction r, such medium is said to be repetitious, or cyclic. The media, cyclic in all directions could be considered but we refer to 1D space as investigated in geophysical theory.

Typical distance between successive local extrema of impedance is of order denoted as $\Delta r$. Consider a medium which is cyclic in a microscale (typical distance $\Delta r$), and when

$$W/\Delta r \rightarrow \infty \, , \, \Delta r > 0 \, , \, W < \infty \tag{6}$$

For global seeing, in macroscale the space metrized as $\rho^*(r^*)$ is applied. The medium appearing in a macroscale is said to be isotropic and pseudohomogeneous if the resulting both

$$\rho(r^*) \approx const. \, , \, \Lambda(r^*) \approx const. \tag{7}$$

pseudoimpedance and contraction coefficient reveal as directionless and constant over a macrospace. As known in geophysical prospecting, the locally alternating parameter exhibits globally the quasi-homogenity while distance is stretched by pseudoanisotropy coefficient, $\Lambda$.

### D. Doppler-like effect

In conformity with scale relationship, equation (4), the scale for homogeneous medium is stretched by a factor $\Lambda$ when passing to pseudohomogeneous medium. The effect of contraction is well known in direct current resistivity sounding over horizontally stratified earth, where $\Lambda$ is termed as pseudoanisotropy coefficient. For horizontally stratified media (proper impedance varying in z-direction only), the governing equation (1) takes the form

$$\frac{\delta^2 U}{\delta x^2} + \frac{\delta^2 U}{\delta y^2} + \frac{\delta^2 U}{\delta z^{*2}} = 0 \tag{8}$$

when asterix refers to alternative as to homogeneity of a medium. The options $\delta z^* = \delta z$ (inhomogeneous medium), or $\delta z^* = \Lambda \delta z$ (pseudohomogeneous medium) yield different depth of a layer. Eventually, but it is not the practice, the horizontal scales are to be contracted, yielding $\delta x^* = \delta x/\Lambda$, and $\delta y^* = \delta y/\Lambda$.

Doppler-likeness of this effect consists in the following. For 3-D pseudohomogeneous medium, we have

$$\delta x^* = \Lambda \delta x, \, \delta y^* = \Lambda \delta y, \, \delta z^* = \Lambda \delta z \tag{9},$$

and substitution $\omega^* = \Lambda \omega$ is possible in the governing equation (1), then becoming

$$\frac{\delta^2 U}{\partial x^{*2}} + \frac{\delta^2 U}{\delta y^{*2}} + \frac{\delta^2 U}{\delta z^{*2}} + \frac{\omega^{*2}}{v^2} U = 0, \, U=U(r^*,t), \, r^* = r^*(x^*,y^*,z^*) \tag{10}.$$

Now anomalous wave frequency $\omega^*$ may take interpretation either in a fine scale of inhomogeneities, equation (1), or in a coarser scale of regional investigation, equation (11). Thus contraction coefficient, $\theta$, appears as a static counterpart of Doppler shift, as detailed further on.

Relative to the expansion, the governing equation could be read

$$\frac{\delta^2 U}{\partial x^2} + \frac{\delta^2 U}{\delta y^2} + \frac{\delta^2 U}{\delta z^2} + \frac{\omega}{v^{*2}} U = 0, \, U=U(r,t), \, v^* = \Lambda v \tag{11}.$$

### E. Discussion: expansion or inhomogeneities?

Introducing equation (5) into (4) yields

$$r^* - r = \theta r \tag{12}$$

which in static systems is quantified by

*distance correction = contraction coefficient • simple distance.*

When the time is introduced to reach kinematic systems, the shift related to equation (11) is quantified by

*emitted frequency – observed frequency = coeff. • observed frequency*

with

*coeff.= (observed wavelength – emitted wanelength) / emitted wavelength.*

Similarly, relative to the expanding universe, the expansion rate related to equation (12) is quantified by

*receding velocity of a galaxy = Hubble's constant • current distant to the galaxy.*

Note that contraction coefficient is dimensionless quantity whereas Hubble's constant is dimensioned in $t^{-1}$.

Let us consider an example of 1D space divided between large segments, $r_v$, with void occupancy individualized with proper impedance $\rho_v$, and small segments, $r_p$, with particle occupancy individualized with proper impedance $\rho_p$. The large and small segments interleave and value of $10^{32}$ is assumed for the ratio $r_v/r_p$. In the 1D universe of such sparsely distributed particles, a very high impedance's contrast is to be assigned for contraction coefficient to match the absolute value of the Hubble's constant.

On the other hand, when there is a number of scales, in hierarchical sequence, the effects of subsequent reparemetrizations are multiplied[2]. Let us take into investigation a mega-region of size much larger than macro-region W. Let the same relations are settled between mega- and macro- scales, as it was for macro- and micro- scales. The outcoming pseudoanisotropy coefficient is obtained as a product of coefficients graded by the size of corresponding subregions. Effective Doppler-like shift is increasing with the size of a region under investigation.

**Conclusions**

The spatial contraction attributed to the pseudohomogeneities (static system) is identified as a counterpart of Doppler shift in a dynamical system. When nothing is known about kinematics of a system both causes might share in the effect. In particular, the contested expansion of universe has an alternative in contraction due to inhomogeneities.

*Acknowledgment.* This research was supported by a grant from the State Committee for Scientific Research (KBN Poland).